\newcommand{\cm}{{~\rm cm}}
\newcommand{\s}{{~\rm s}}
\newcommand{\km}{{~\rm km}}
\newcommand{\g}{{~\rm g}}
\newcommand{\K}{{~\rm K}}
\newcommand{\erg}{{~\rm erg}}
\newcommand{\yr}{{~\rm yr}}
\newcommand{\Myr}{{~\rm Myr}}

\newcommand{\kpc}{{~\rm kpc}}

\newcommand{\dyn}{{~\rm dyn}}

\documentclass[12pt,preprint]{aastex}
\begin{document}

\title{SOUND WAVES EXCITATION BY JET-INFLATED BUBBLES IN CLUSTERS OF GALAXIES}

\author{Assaf Sternberg\altaffilmark{1},  Noam Soker\altaffilmark{1}}

\altaffiltext{1}{Department of Physics,
Technion$-$Israel Institute of Technology, Haifa 32000, Israel;
phassaf@techunix.technion.ac.il; soker@physics.technion.ac.il}

\begin{abstract}
We show that repeated sound waves in the intracluster medium (ICM) can be
excited by a single inflation episode of an opposite bubble pair.
To reproduce this behavior in numerical simulations the bubbles should be inflated by jets,
rather than being injected artificially.
The multiple sound waves are excited by the motion of the
bubble-ICM boundary that is caused by vortices inside the inflated bubbles
and the backflow (`cocoon') of the ICM around the bubble.
These sound waves form a structure that can account for the ripples observed
in the Perseus cooling flow cluster. We inflate the bubbles using slow massive
jets, with either a wide opening angle or that are precessing.
The jets are slow in the sense that they are highly sub-relativistic, $v_j \sim 0.01c-0.1c$,
and they are massive in the sense that the pair of bubbles carry back to the
ICM a large fraction of the cooling mass, i.e., $\sim 1-50 M_\odot \yr^{-1}$.
We use a two-dimensional axisymmetric (referred to as 2.5D) hydrodynamical
numerical code (VH-1).
\end{abstract}

{{\it Subject headings:} (galaxies:) cooling flows galaxies: clusters:
 general galaxies: jets}

\section{INTRODUCTION}
\label{sec:intro}

Higher X-ray emissivity arcs, termed ripples, are observed in the
intracluster medium (ICM) of the Perseus cluster (Fabian et al. 2003, 2006;
Sanders \& Fabian 2007), as well as in A2052 (Blanton et al. 2007).
The ripples are thought to be sound waves that are
excited by the inflation of bubbles near the cluster's center.
Some of the ripples are observed on the outer edge of bubbles; they are compressed ICM
that will turn into shock waves, and later to sound waves.
It is possible that some of the other
ripples are compressed by weak jets (Soker \& Pizzolato 2005), rather than
being sound waves detached from their parent bubble. Beside being interesting
by their own merit, the sound waves can heat the ICM and partially offset the
radiative cooling of the ICM (e.g., Br\"uggen et al. 2005; Fabian et al.
2005; Fujita \& Suzuki 2005; Graham et al. 2008). \par

To obtain repeated sound waves previous numerical simulations assumed
repeating bubble inflation episodes (e.g., Ruszkowski et al. 2004a,b;
Sijacki \& Springel 2006a, b). This had to be done because these
simulations used \emph{artificial bubbles}, i.e., numerically injected
spherical bubbles at off-center locations, as is commonly done (e.g.,
Br\"uggen 2003; Br\"uggen \& Kaiser 2001; Gardini 2007; Jones \& De Young
2005; Pavlovski et al. 2008; Reynolds et al. 2005; Robinson et al. 2004;
Ruszkowski et al. 2004a, 2004b, 2007, 2008; Scannapieco \& Br\"uggen 2008).
This of course is not the way bubbles are formed in clusters. Bubble are
formed by jets [a different approach of injecting magnetic energy instead
of a jet (Nakamura et al. 2008; Xu et al. 2008) requires further study].
When large, almost spherical bubbles (termed \emph{fat bubbles}) are inflated
by jets, some phenomena appear, that are not revealed when artificial bubbles
are used (Sternberg \& Soker 2008b). Among these
are: (1) more stable bubbles, and (2) bubble-ICM boundary that is
stochastically vibrates due to the vortices inside the bubble and the back
flowing ICM.

In previous works we found that in order for the jets to inflate fat bubbles
they should have a large opening angle, or they should rapidly precess (Soker
2004, 2006; Sternberg et al. 2007; Sternberg \& Soker 2008a).
In addition, jets that are relatively slow,
$v_j \sim 10^4 \km \s^{-1} \ll c$,  and with a mass loss rate of the order of
$2\dot M_j \sim1-50 M_\odot \yr^{-1}$ (for both jets), are more likely to
inflate fat bubbles.
The same results can be seen in the simulations of Omma et al. (2004), Alouani Bibi et al. (2007)
and Binney et al. (2007), who use similar parameters to ours, but instead of conical jets with
wide opening angles they use cylindrical jets with large initial radius.
This is also shown by the simulation of Heinz et al. (2006), who manages to
inflate a bubble (although not a bubble attached to the center) by launching
a jet with a
velocity of $3\times 10^4 \km \s^{-1}$ and a mass loss rate of
$35 M_\odot \yr^{-1}$ in one jet.
In this paper we continue our study of jet-inflated
bubbles, and show that such bubbles can excite several consecutive sound
waves without the need to invoke periodic jet launching episodes. We do
not deal with the propagation of sound waves and their properties. These
seem to require more sophisticated treatment (Graham et al. 2008). We limit
ourself to show that a proper treatment of bubble inflation can better
explain the excitation of sound waves in the ICM. \par

\section{NUMERICAL METHOD AND SETUP}
\label{sec:numerics}

The simulations were performed using the \emph{Virginia Hydrodynamics-I}
code (VH-1; Blondin et al. 1990; Stevens et al. 1992), as described in
Sternberg \& Soker (2008b). In this paper we mention only the important
features of the code. The unperturbed ICM temperature is set to
$2.7 \times 10^7 \K$. Gravity is included, but radiative cooling is not
included. We study a three-dimensional axisymmetric flow with a 2D grid
(referred to as 2.5D). We simulate half of the meridional plane
using the two-dimensional version of the code in spherical coordinates.
The symmetry axis of all plots shown in this paper is along the x (horizontal)
axis. \par

We run two cases. In the wide jets case, the two opposite jets were injected
at a radius of $0.1 \kpc$, with constant mass flux of
$\dot M_j= 5 M_\odot \yr^{-1}$ per one jet, and a constant radial velocity of
$v_j= 7750 \km \s^{-1}$, inside a half opening angle of $\alpha = 70^\circ$.
The total power of the two jets is $\dot E_{2j}= 2\times10^{44} \erg \s^{-1}$.
The jets were active for a period of $\Delta t_j= 10 \Myr$,
from $t=-10 \Myr$ until $t=0$. For more detail on the numerical code and the
properties of the bubbles see Sternberg \& Soker (2008b). \par

In the case of the precessing jets the two opposite jets were injected
at a radius of $0.1 \kpc$ with $\dot M_j= 5 M_\odot \yr^{-1}$ per one jet,
and a constant radial velocity of $v_j= 7750 \km \s^{-1}$
(the same parameters of the wide jets), but
within a half opening angle of $\alpha=5^\circ$. In our axisymmetric code
the jets are precessing very rapidly around the symmetry axis (i.e., in 3D
we actually inject a torus). We vary the angle between the symmetry axis and
the jets' axis $\theta$ in a random way. Namely, the jet axis has the same
probability to take any direction within an angle of
$5^\circ\geq\theta\leq45^\circ $. This is done by changing $\theta$
periodically and taking $d(\cos\theta)/dt$ to be constant. The precession
period, i.e., the time the jet returns to the same angle $\theta$, is
$T_{\rm prec}=0.1 \Myr$. The jet's interaction with the ICM is similar to that
of a wide jet with a half opening angle of $\alpha\sim 50^\circ$. The jets
were active for a duration of $\Delta t_j= 18 \Myr$ between $t=-18\Myr$ and
$t=0$. Gravity was included as was in the wide jets case.

\section{RESULTS}
\label{sec:results}

\subsection{Wide jets}
\label{sec:wide}
\begin{figure}
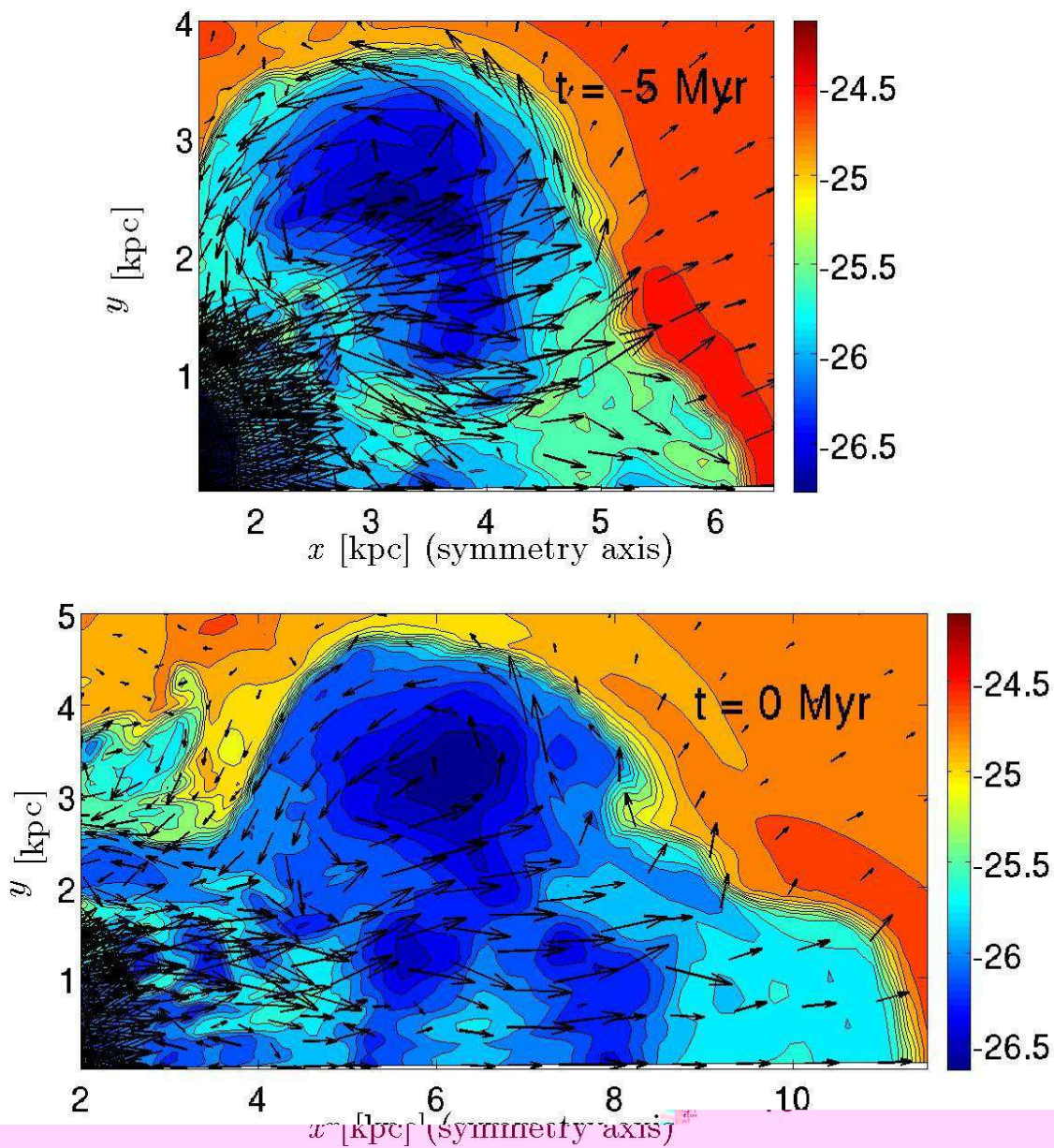
  
\begin{tabular}{c}
\hskip -1.0 cm
{\includegraphics[scale=0.5]{vortex_-5Myr.eps2}} \\
\hskip -1.0 cm
{\includegraphics[scale=0.5]{vortex_0Myr.eps2}}  
\end{tabular}
\caption{Density and velocity map for the wide-jet-inflated bubbles at
$t=-5~$Myr, and at $t=0$, when the jet is shutoff (the jet was turned on
at $t=-10~$Myr). The density scale is on the right in logarithmic scale and
cgs units. The arrows represent the velocity of the flow:
$0.1C_s <v_j \leq 0.5C_s$ (shortest), $0.5C_s<v_j \leq C_s$,
$C_s<v_j \leq 5C_s$, and $5C_s<v_j \leq 10C_s$ (longest in this case).
Here $C_s=775 \km \s^{-1}$ is the sound speed of the undisturbed ICM.
Note the different length scale in the two panels. }
\label{fig:jeti}
\end{figure}

In Fig. \ref{fig:jeti} we show the density and velocity map of the
wide-jet-inflated bubble at $t=-5~$Myr, and at $t=0$ when the jet is shut off.
There are some flow structures which characterize the jet-inflated bubble
that are not present when artificial bubbles are used (Sternberg \&
Soker 2008b). Most relevant for us are the following: (1) The bubble and the
dense shell around it have radial momentum. In particular, the shell's front
is relatively dense and has an outward velocity. This will form the ripple
observed at the front edge of the bubbles in Perseus.
(2) As the bubbles grow their shapes change. This relative motion of the
bubble-ICM boundary excites sound waves.
(3) There is a circular flow, a vortex, around the lowest density region
in the bubble. This vortex and the back flow of the ICM along the boundary
of the bubble change the shape of the bubble, a processes leading to
excitation of sound waves. We also note the back flow of the ICM around the
bubble, and that at early times the bubble expands supersonically, e.g., an
average Mach number of $\sim 1.3$ between $t=-5~$Myr and $t=0$, and the
shell is a weak shock. At later times the bubble moves subsonically. \par

In Fig. \ref{fig:sound1} we omit low densities (i.e., the bubble material)
and show only the high density regions of the ICM. Therefore, the bubble
interior is white. The density scale is logarithmic scale of cgs units (i.e.,
$\g \cm^{-3}$), emphasizes the small density variations resulting from the
sound waves and weak shocks excited by the bubble. Several dense arcs are seen
in the ICM as they expand outward. We note the dense arc at the right-bubble
front at times $t=-5$, $0$, and $10~$Myr, but at $t=5~$Myr the density
close to the bubble front, at $(x,y) \simeq (14.5,0) \kpc$, is relatively low.
The arc is compressed, and sends a sound wave outward. At $t=5~$Myr the
region at the right-bubble front is near the rarefaction phase of the wave,
the trough of the wave.
The change in the shape of the bubbles is evident also from comparing the
two opposite bubbles. \par

\begin{figure}  
\hskip -2.0 cm
\begin{tabular}{cc}
\hskip -0.5 cm
{\includegraphics[scale=0.29]{soundwave_-5Myr.eps2}} &
\hskip -0.7 cm
{\includegraphics[scale=0.29]{soundwave_0Myr.eps2}} \\
\hskip -0.5 cm
{\includegraphics[scale=0.29]{soundwave_5Myr_cuts.eps2}} &
\hskip -0.7 cm
{\includegraphics[scale=0.29]{soundwave_10Myr_cuts.eps2}}
\end{tabular}
\caption{Density map for the ICM around the wide-jet-inflated bubbles at four
times as indicated. The density scale is on the right in logarithmic scale
and cgs units. Several ripples are formed by one episode of bubble pair
production and are clearly seen in these figures. The black solid, dashed,
and dot-dashed lines indicate, respectively, the direction of $\beta= 10^\circ$,
$40^\circ$,and $170^\circ$ in respect to the positive direction of the
$x$-axis. The jets are active from $t=-10~$Myr to $t=0$. }
\label{fig:sound1}
\end{figure}

The main morphological feature we obtain is the presence of several ripples
along each radial direction. The outermost ripple is a weak shock, while the
ripples closer to the bubbles are the crests of sound waves.
The sound waves are excited by the corrugated boundary of the bubble that is
formed mainly by the vortices and the back flow of the ICM. Considering that
two pairs of large bubbles are observed in Perseus (the actual bubble
morphology is more complicated; Dunn et al. 2006), we conclude that the
ripples in the Perseus cluster can be explained without the need to
assume more episodes of jet launching than the number of bubble pairs
that are actually observed. \par

To better explore the properties of the waves we show the density, pressure,
and temperature along three radial cuts at angles of $10^\circ$, $40^\circ$
and $170^\circ$, in respect to the positive direction of the $x$-axis, at
two times, $t=5~$Myr and $t=10~$Myr (corresponding to the lines indicated in
the two late panels in \ref{fig:sound1}). The typical pressure variations
from the average, along these radial cuts, are $\sim 5-20 \%$. Although the 
small perturbations are somewhat lower then those observed in the Perseus 
cluster (Fabian et al. 2006), the stronger perturbations are in good agreement 
with these observations.
The power of our two jets is $\dot E_{2j}= 2\times10^{44} \erg \s^{-1}$, similar
to that of the cavities in Perseus (Rafferty et al. 2006). However, the total
energy injected here to the two bubbles is $\sim 6 \times 10^{58} \erg$,
while the total energy of the bubbles in Perseus is
$\sim 20 \times 10^{58} \erg$ (Rafferty et al. 2006). The density in the
central region of $\sim 10^{-25} \g \cm^{-3}$ is similar to that in Perseus
(Fabian et al. 2003). As we did not try to fit the bubbles of Perseus, we
do not expect a perfect match. Varying the jets properties can lead to a
better fit. In this paper we limit ourself to presenting the basic
physics of the process. In any case, the precessing jets case studied here
form sound waves with larger amplitudes (see below). \par

\begin{figure}  
\begin{tabular}{cc}
\hskip -2.0 cm
{\includegraphics[scale=0.29]{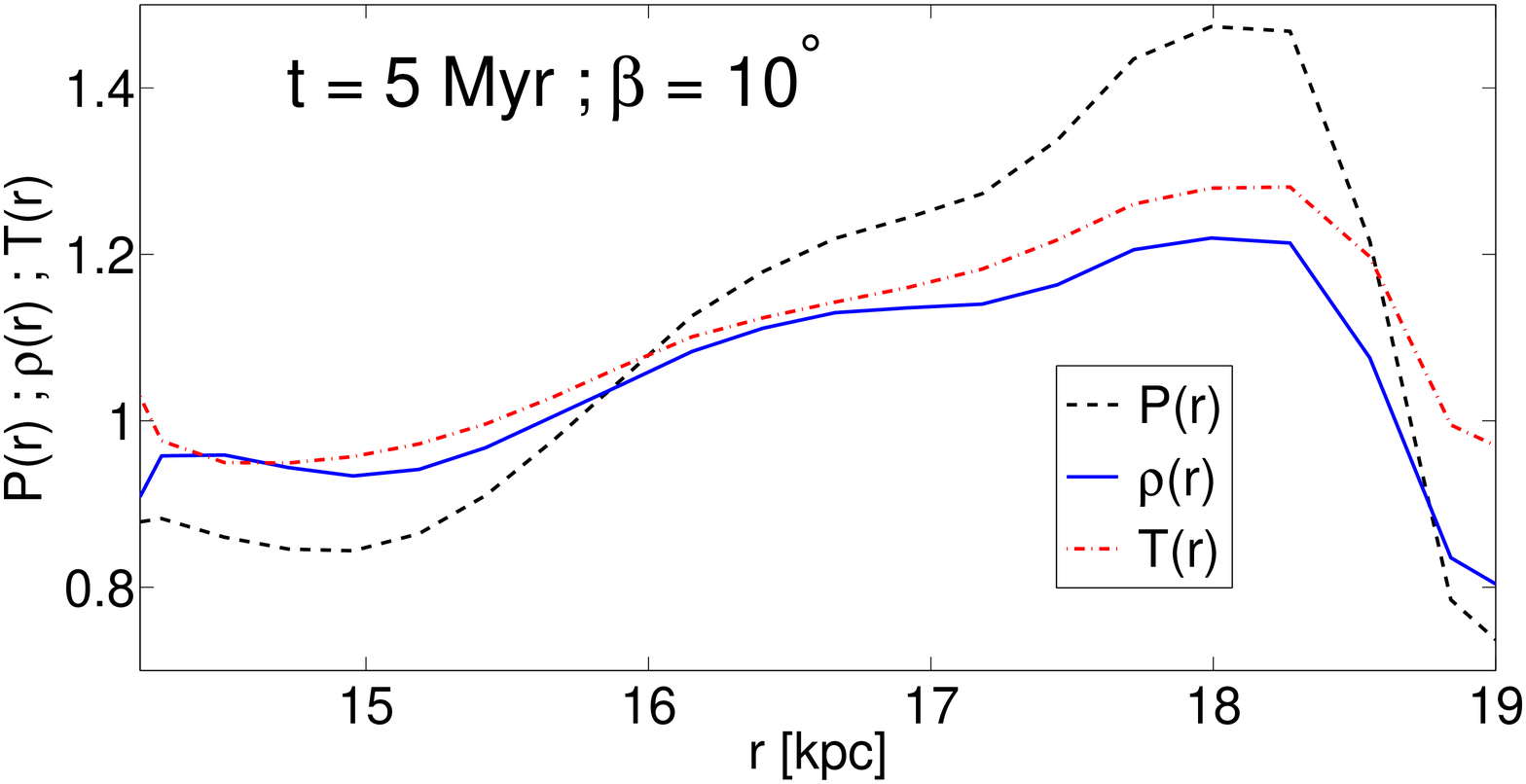}} &
\hskip -1.2 cm
{\includegraphics[scale=0.29]{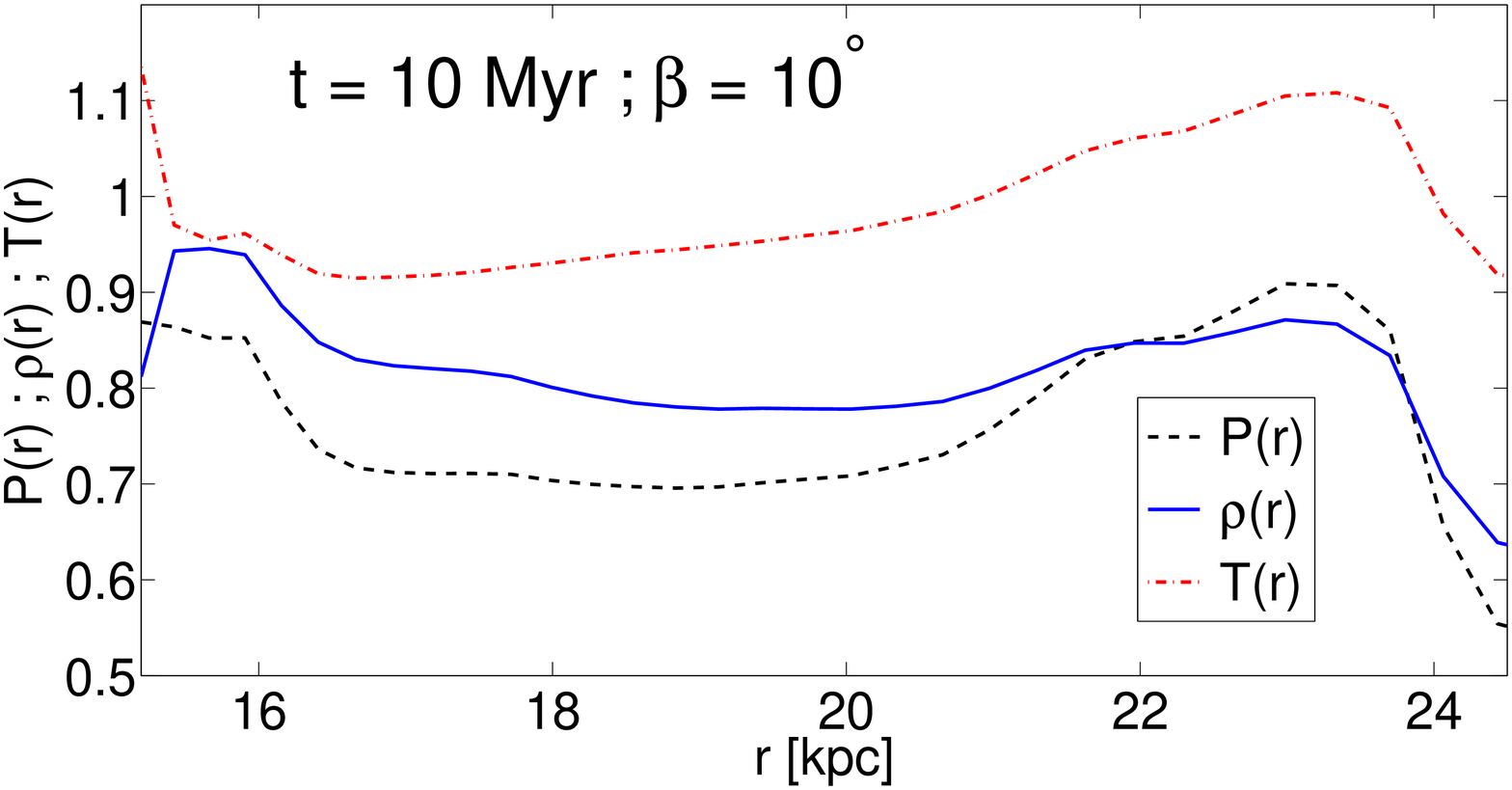}} \\
\hskip -2.0 cm
{\includegraphics[scale=0.29]{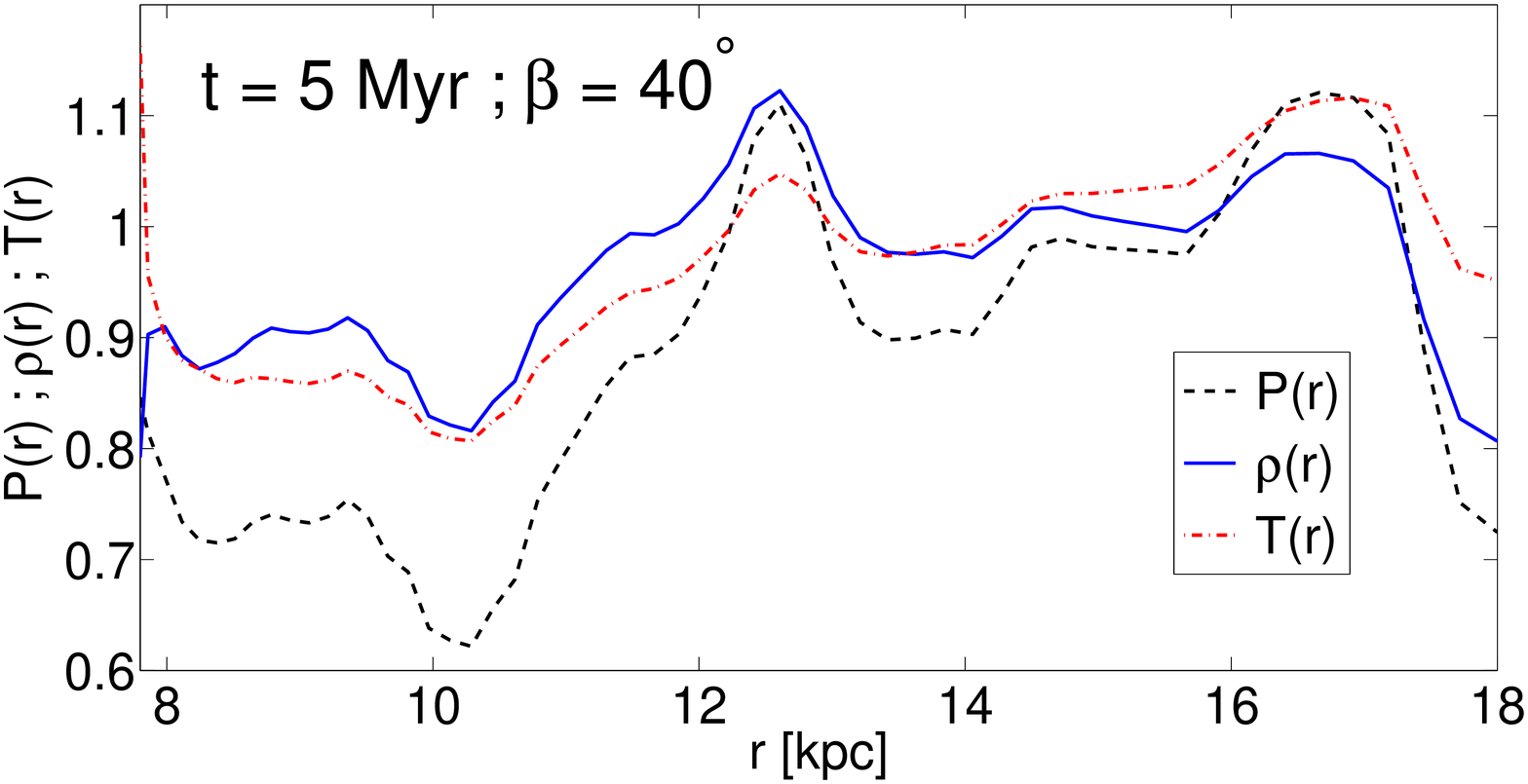}} &
\hskip -1.2 cm
{\includegraphics[scale=0.29]{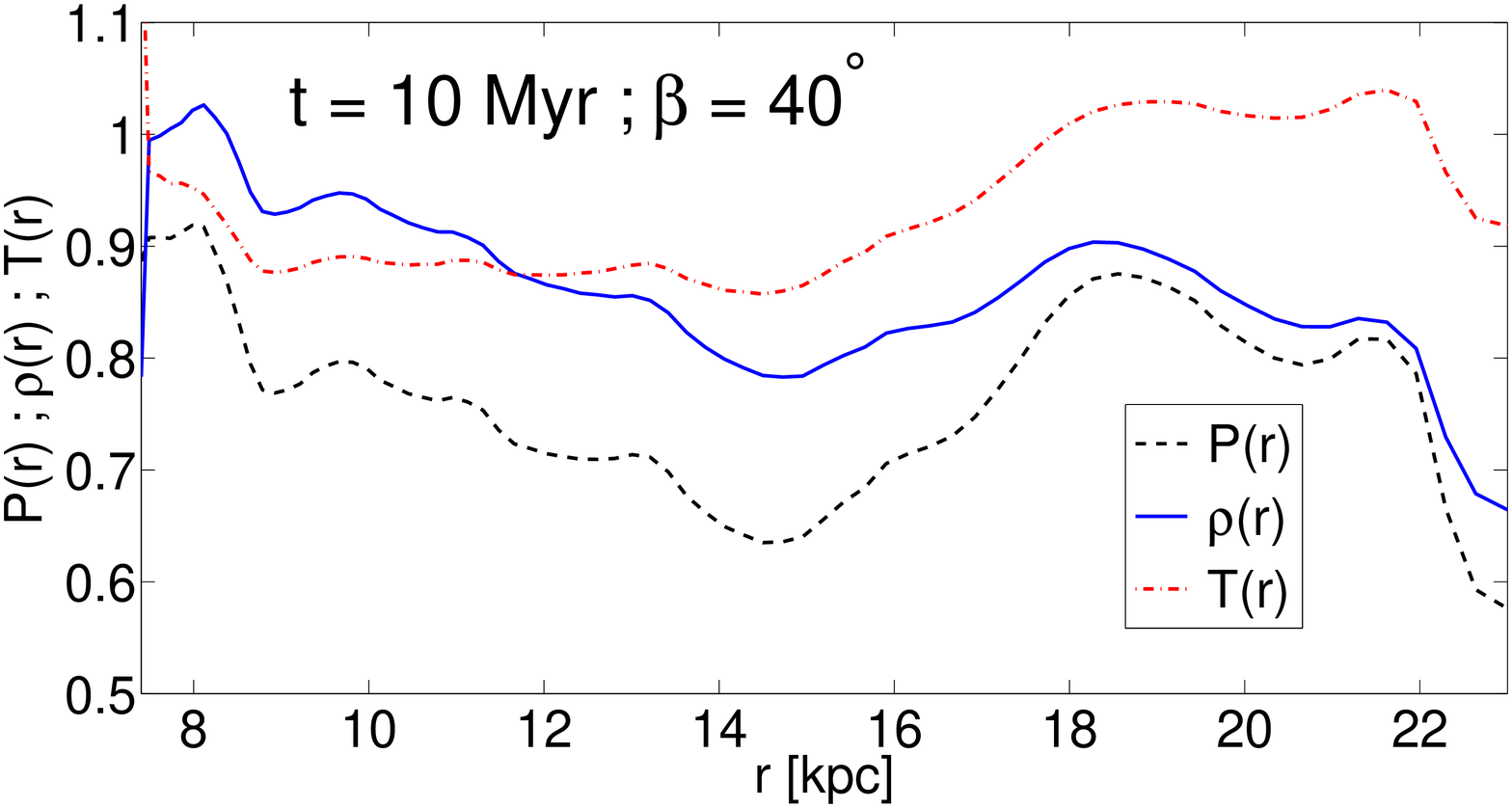}} \\
\hskip -2.0 cm
{\includegraphics[scale=0.29]{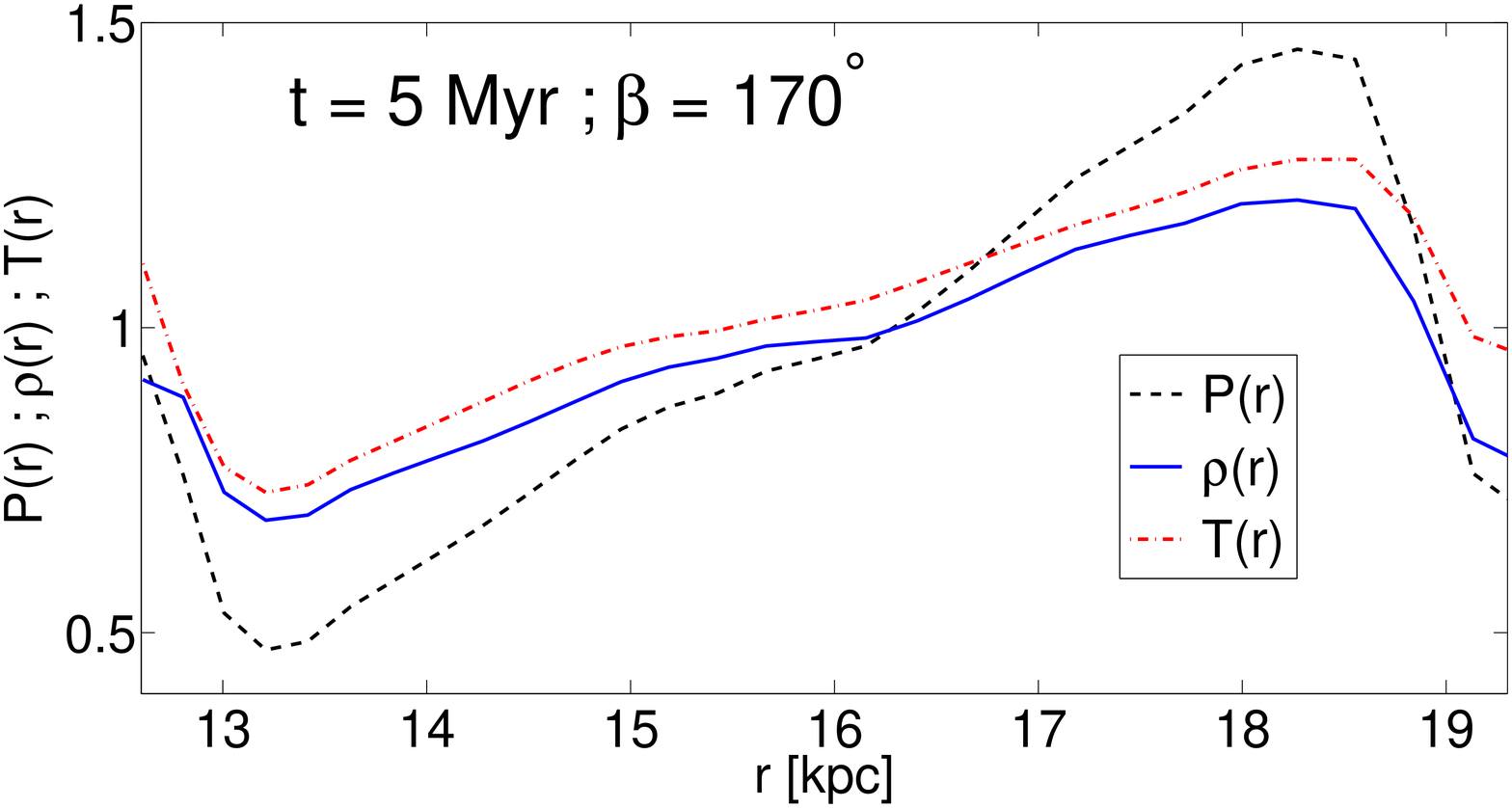}} &
\hskip -1.2 cm
{\includegraphics[scale=0.29]{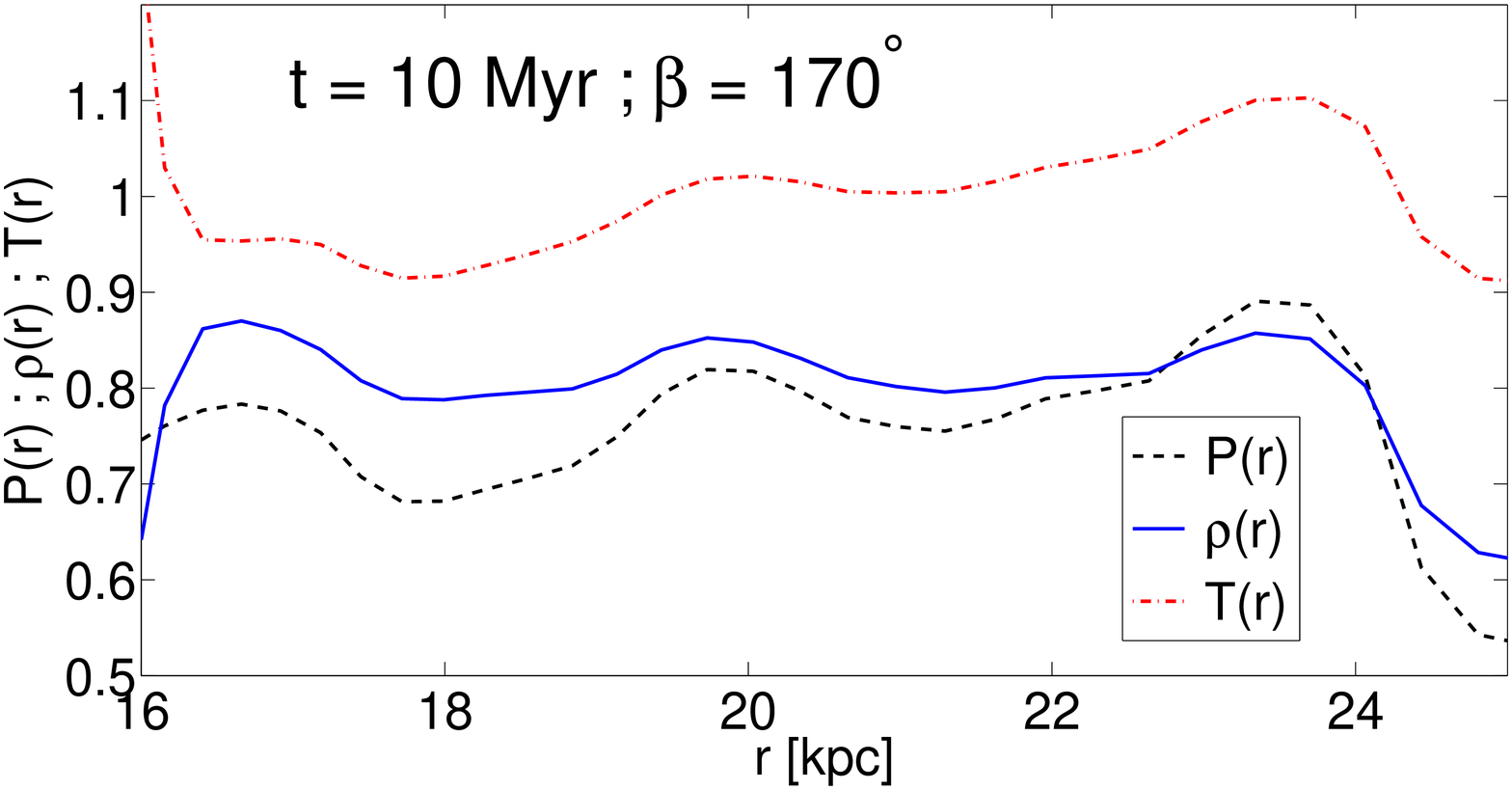}}
\end{tabular}
\caption{Density, temperature, and pressure along three radial cuts at two
epochs, as marked on the last two panels of Fig. \ref{fig:sound1}.
In the left and right panels quantities are taken at $t=5~$Myr and $t=10~$Myr,
respectively. The radial directions are marked on the panels.
The pressure is give in units of $6\times10^{-10} \dyn \cm^{-2}$,
density in units of $1.5\times10^{-25} \g \cm^{-3}$, and temperature in units 
of $2.7\times10^7 \K$.}
\label{fig:sound2}
\end{figure}

\subsection{Precessing jets}
\label{sec:precessing}

In Fig. \ref{fig:sound1p} we omit low densities (i.e., the bubble material),
and show only the high density regions of the ICM. Therefore, the bubble
interior is white. The density scale is logarithmic scale of cgs units
(i.e., $\g \cm^{-3}$). As with the wide jets, several ripples are seen in
the ICM as they expand outward. The precessing jets result in a much faster
variations in the shape of the bubble-ICM boundary, and in different regions
at different times. Therefore, the ripples are more pronounced. \par

\begin{figure}  
\hskip -2.0 cm
\begin{tabular}{cc}
\hskip -0.5 cm
{\includegraphics[scale=0.29]{soundwavep_-10Myr.eps2}} &
\hskip -0.7 cm
{\includegraphics[scale=0.29]{soundwavep_-5Myr.eps2}} \\
\hskip -0.5 cm
{\includegraphics[scale=0.29]{soundwavep_0Myr_cuts.eps2}} &
\hskip -0.7 cm
{\includegraphics[scale=0.29]{soundwavep_5Myr_cuts.eps2}}
%
%
\end{tabular}
\caption{
Density map for the ICM around the precessing-jet-inflated bubbles at four
times as indicated. The density scale is on the right in logarithmic scale
and cgs units. Several ripples are formed by one episode of bubble pair
production and are clearly seen in these figures. The black solid, dashed,
and dot-dashed lines indicate, respectively, the direction of
$\beta= 10^\circ$, $40^\circ$,and $80^\circ$ in respect to the positive
direction of the $x$-axis. The jets were active from $t=-18~$Myr to $t=0$.}
\label{fig:sound1p}
\end{figure}

\begin{figure}  
\begin{tabular}{cc}
\hskip -2.0 cm
{\includegraphics[scale=0.29]{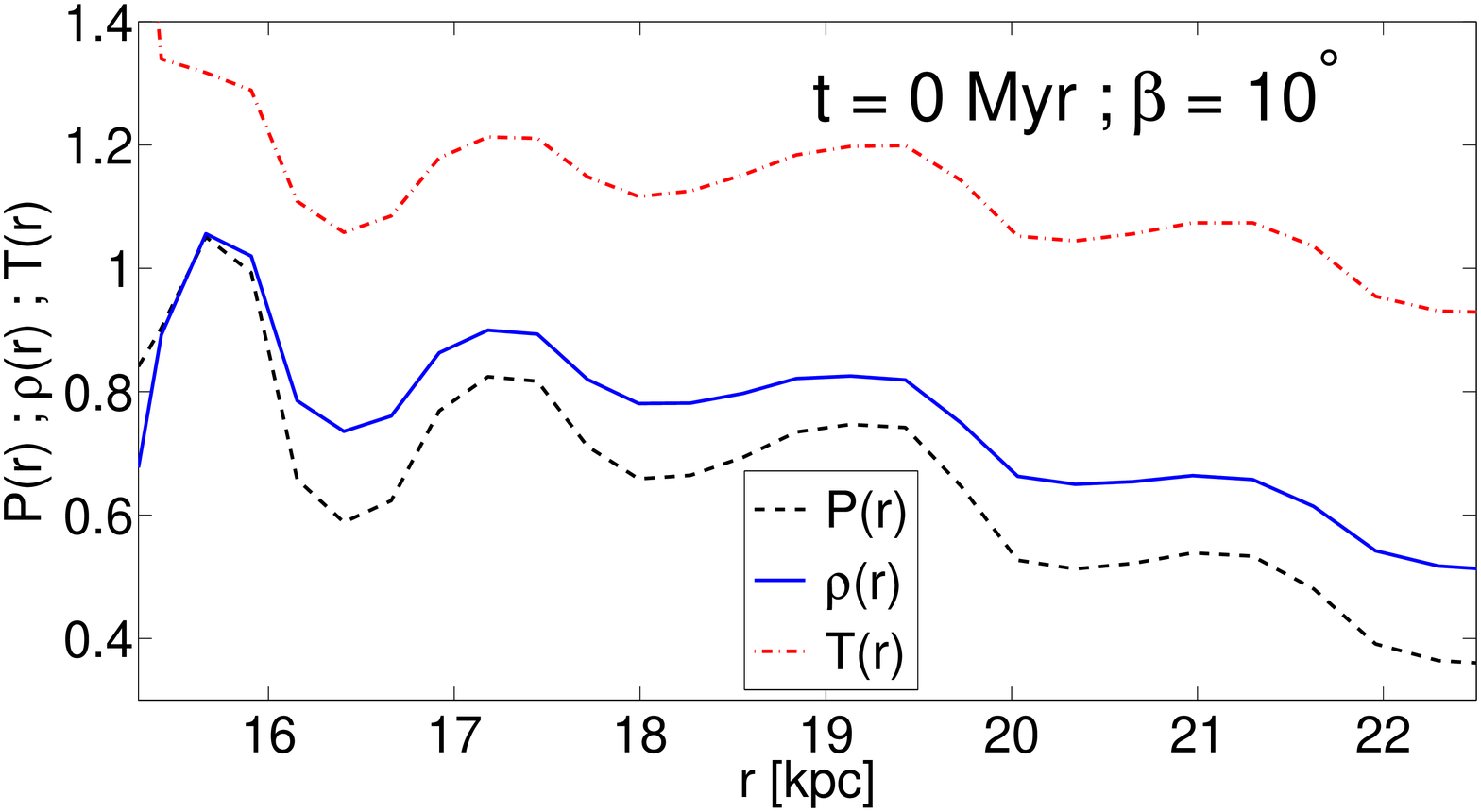}} &
\hskip -1.2 cm
{\includegraphics[scale=0.29]{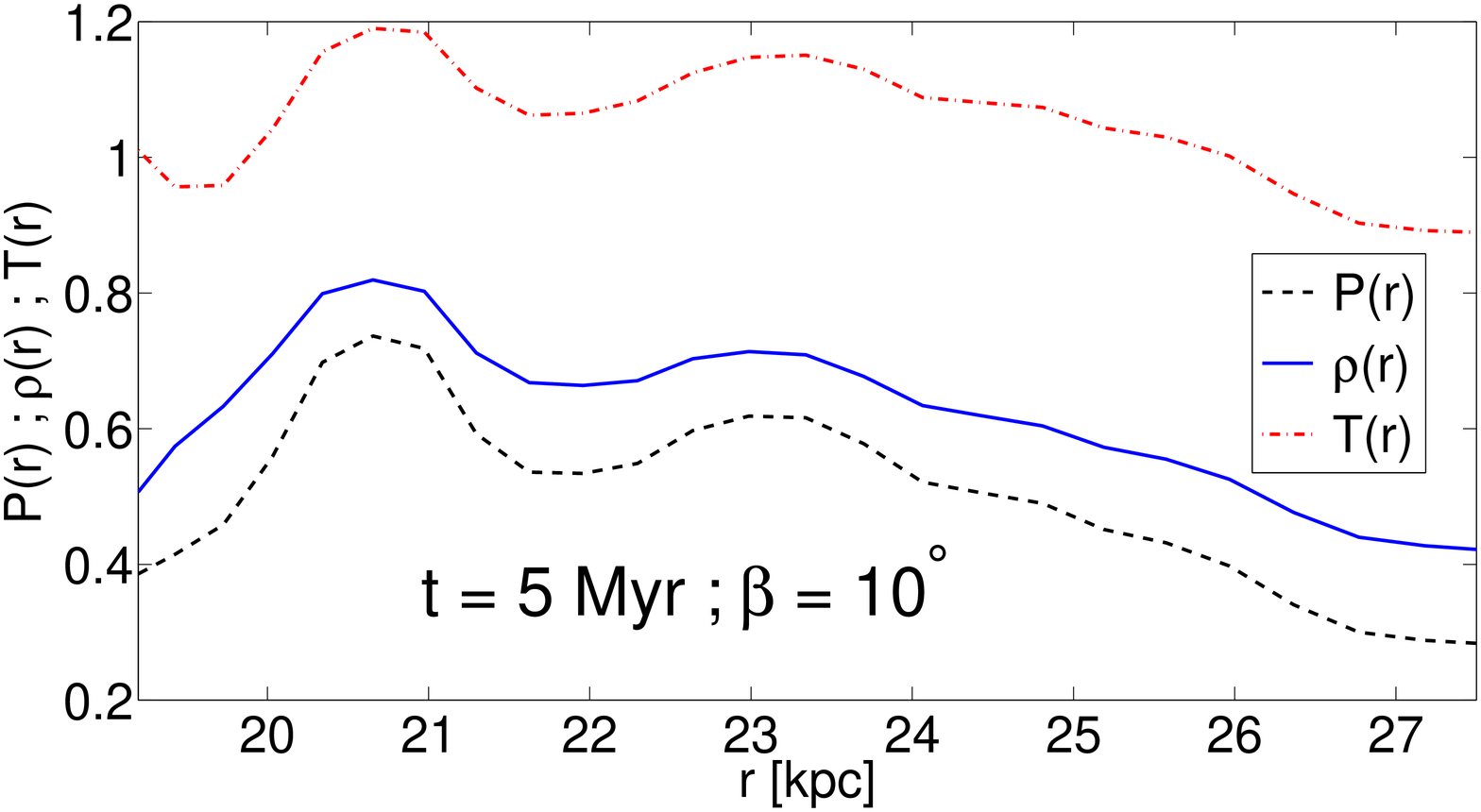}} \\
\hskip -2.0 cm
{\includegraphics[scale=0.29]{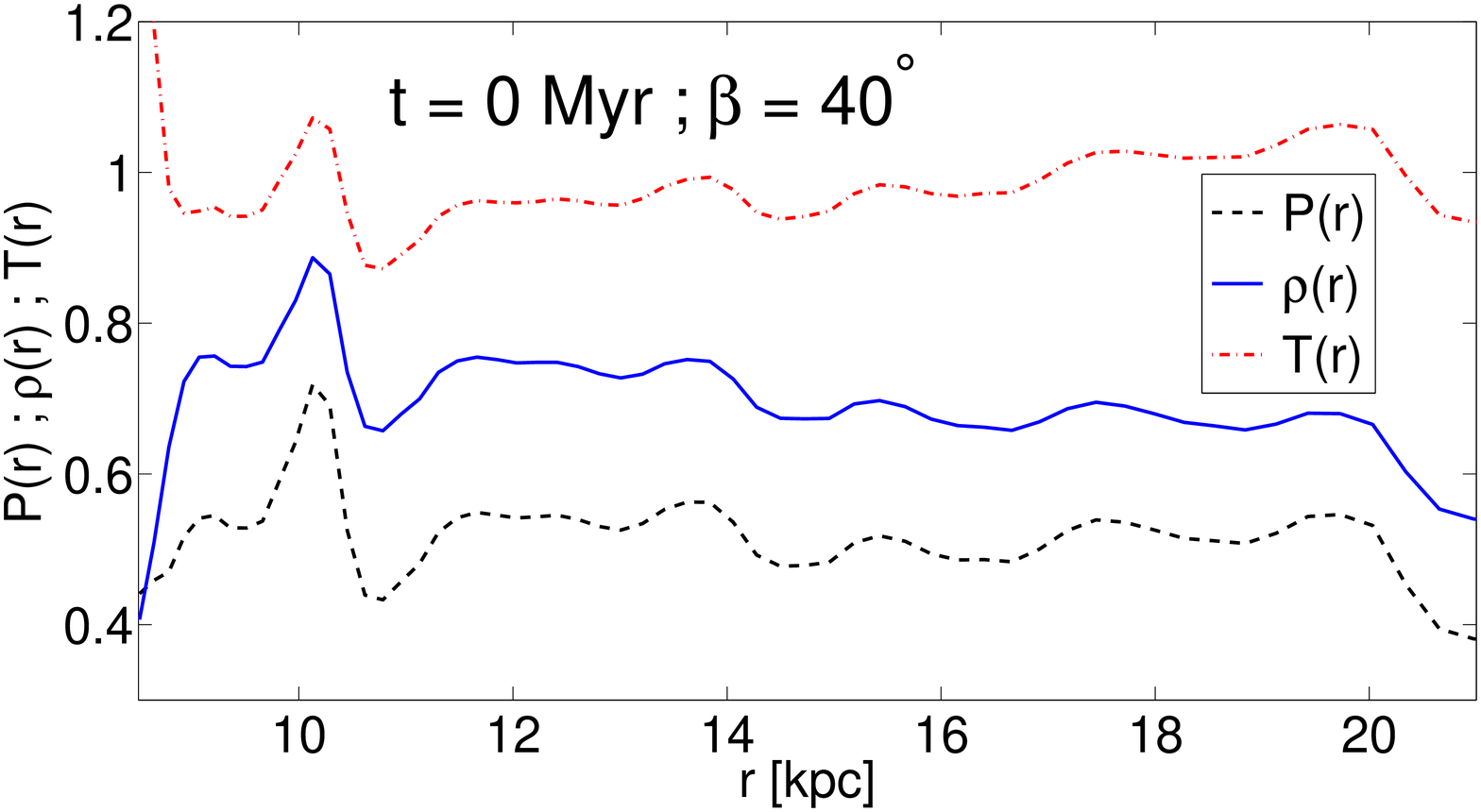}} &
\hskip -1.2 cm
{\includegraphics[scale=0.29]{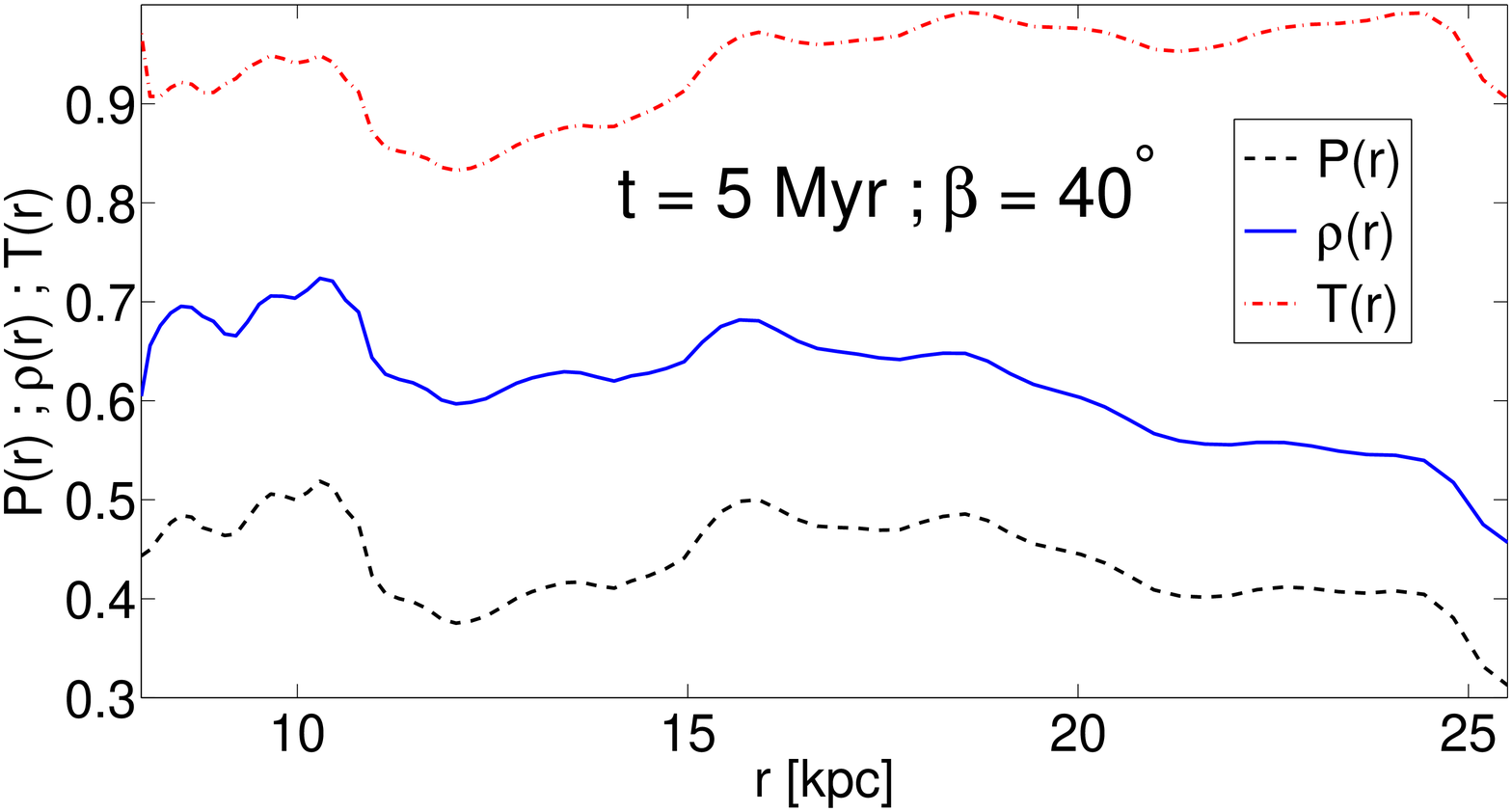}} \\
\hskip -2.0 cm
{\includegraphics[scale=0.29]{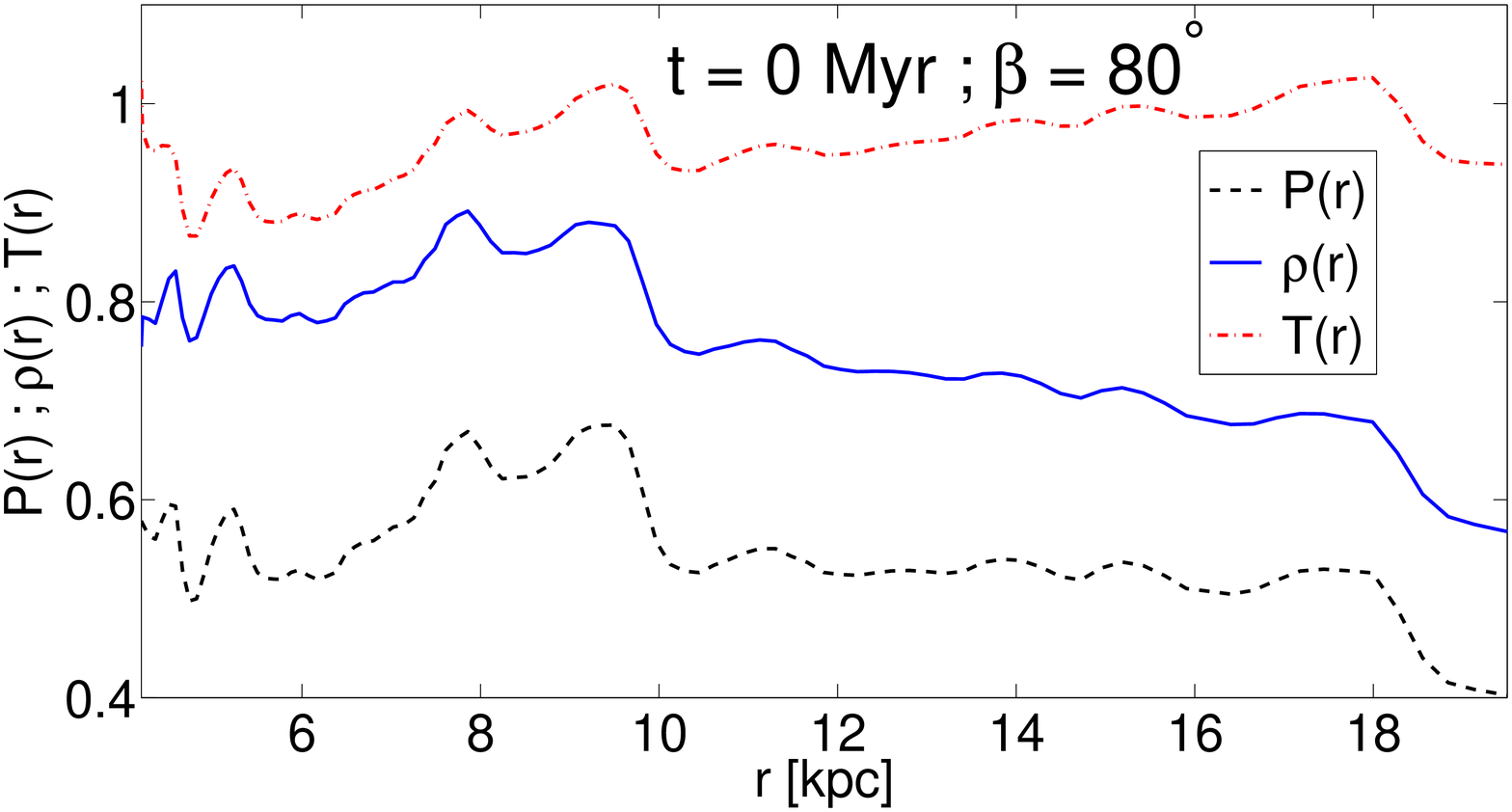}} &
\hskip -1.2 cm
{\includegraphics[scale=0.29]{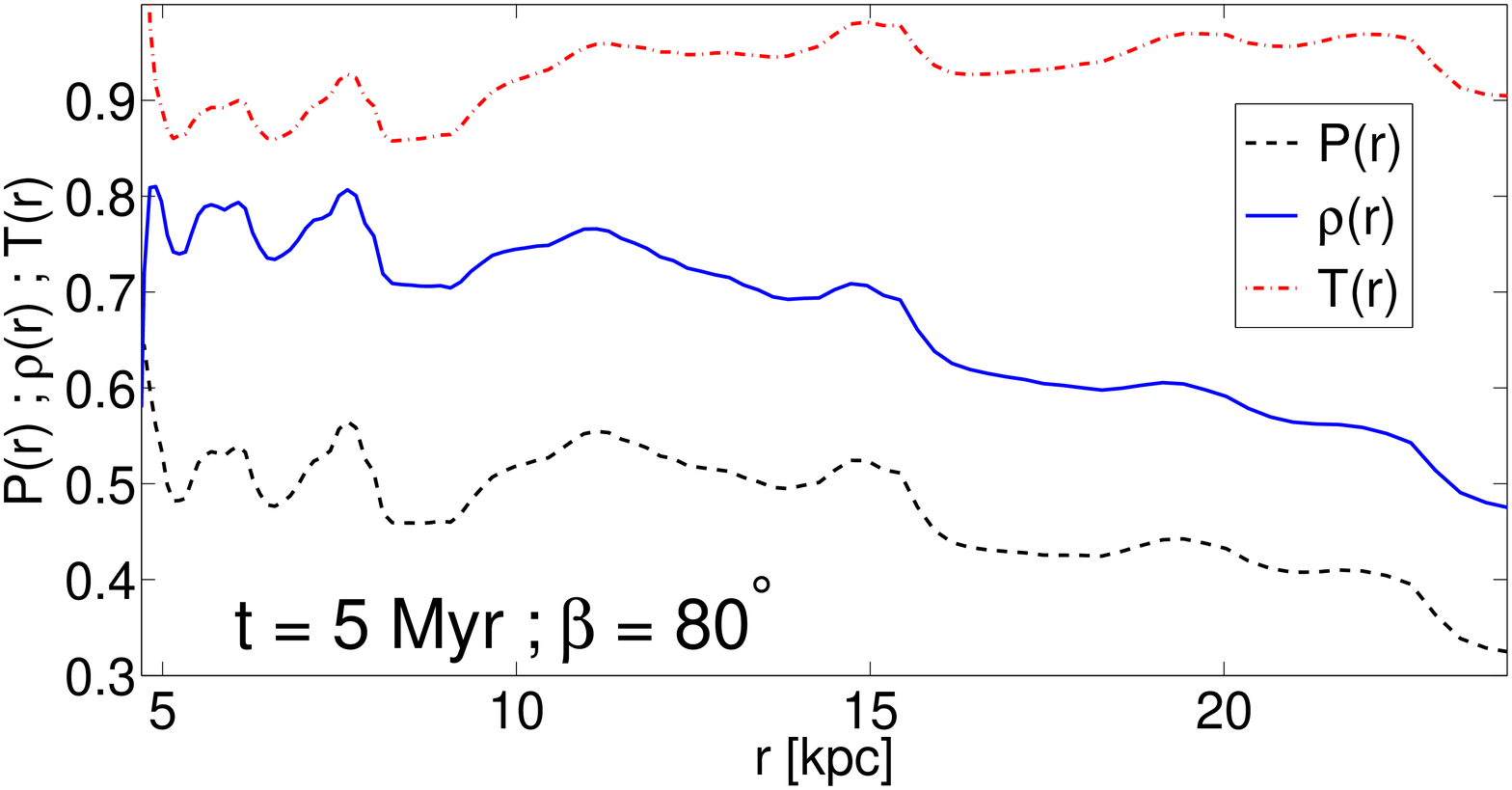}}
\end{tabular}
\caption{Density, temperature, and pressure along three radial cuts at two
epochs, as marked on the last two panels of Fig. \ref{fig:sound1p}.
In the left and right panels quantities are taken at $t=0~$Myr and $t=5~$Myr,
respectively. The radial directions are marked on the panels.
The pressure is give in units of $9\times10^{-9} \dyn \cm^{-2}$,
density in units of $2\times10^{-25} \g \cm^{-3}$, and temperature in units of
$2.7\times10^7 \K$.}
\label{fig:sound2p}
\end{figure}

In figure \ref{fig:sound2p} we show the density, pressure, and temperature
along three radial cuts at angles of $10^\circ$, $40^\circ$ and $80^\circ$,
in respect to the positive direction of the $x$-axis, at two times, $t=0~$Myr
and $t=5~$Myr (corresponding to the lines indicated in
the two late panels in \ref{fig:sound1p}). The typical pressure variations 
from the average, along these radial cuts, are $\sim 10-20 \%$, making them 
a somewhat better fit to the ripples observed in the Perseus cluster, than 
the ripples in the wide jets case.

\section{DISCUSSION AND SUMMARY}
\label{sec:discussion}

We followed the response of the ICM to the inflation of bubbles by jets.
The inflation of a bubble by a jet results in vortices inside the bubble and
a backflow of the ICM around the bubble (Sternberg et al. 2007; Sternberg \&
Soker 2008a, b). Both processes cause the bubble-ICM to change shape with
time and to become corrugated. In the case of a precessing jet (section
\ref{sec:precessing}), the changes in the jet axis cause a more prominent
change in the bubble-ICM shape.
This motion of the bubble-ICM boundary sends shocks and sound waves into the ICM.
This effect cannot be reproduced by artificial bubbles,
i.e., by numerically inserting spherical bubbles at off-center locations. \par

Our main result is that one episode of a bubble pair inflation can excite
several sound waves along each radial direction. These form high-density
arcs, the ripples. The front ripple is actually a weak shock. There is no
need to introduce periodic (or semi-periodic) episodes of bubble inflation.
Nonetheless, multiple episodes will make more ripples, and make the ripple
structure more complicated.
This is the case in Perseus, where the main central bubble
pair was formed by jets that changed their direction and intensity
(Dunn et al. 2006).

We also observe a dense arc at the bubble's front that is apparent part of
the time. The dense arc is the excitation of a new sound wave at the
compression phase. Part of the time the region at the bubble's front is at
the rarefaction phase, the wave trough, and no dense arc is observed there. \par

Our results can be incorporated into a broader scope. In a previous paper
(Sternberg \& Soker 2008b) we found that to follow the evolution of bubbles
in the ICM one must inflate them by jets, rather than introduce them artificially.
The evolution of bubbles and their influence on the ICM, e.g.,
sound waves, is crucial for the deposition of heat from the central active
galactic nuclei to the ICM. Our works show that in order to understand the
heating of the ICM one must properly inflate bubbles. \par

\acknowledgements
We thank John Blondin for his immense help with the numerical code. This
research was supported by the Asher Fund for Space Research at the Technion,
and by the Israeli Science Foundation (grant No. 89/08).

\end{document}